\documentclass[twocolumn,showpacs,amsmath,amssymb,floatfix,prx,superscriptaddress]{revtex4-1}
\usepackage[latin1]{inputenc} 
\usepackage{setspace}
\usepackage{graphicx}
\usepackage{graphics}
\usepackage{subfigure}
\usepackage{dcolumn}
\usepackage{color}
\usepackage{SIunits}
\usepackage{mathtools}
\usepackage{amsmath, amssymb, bm}
\usepackage[normalem]{ulem}
\usepackage{url}
\usepackage[left, modulo, pagewise]{lineno}
\usepackage{afterpage}
\usepackage{wrapfig}
\usepackage{enumitem}
\usepackage{romannum}
\usepackage{array,xcolor,colortbl}
\newcolumntype{G}{>{\centering}p{0.2\textwidth}}
\newcolumntype{C}{>{\centering\arraybackslash}p{0.065\textwidth}}
\usepackage[T1]{fontenc}

\usepackage{array}
\usepackage{multirow}
\usepackage{indentfirst}
\usepackage{makecell}

\newcommand{\refpanel}[2]{\hyperref[#1]{\ref*{#1}--\textsf{#2}}}
\newcommand{\equaref}[1]{\hyperref[#1]{\!(\ref*{#1})}}

\usepackage{hyperref}
\hypersetup{
    colorlinks,
    urlcolor={black},
    linkcolor={black},
    citecolor={black}}

\begin{document}

\title{How to generate intense isolated attosecond pulses from relativistic plasma mirrors?}
\author{H. Kallala} 
\affiliation{LIDYL, CEA, CNRS, Universit\'e Paris-Saclay, CEA Saclay, 91 191 Gif-sur-Yvette, France\\}
\affiliation{Maison de la Simulation, CEA, CNRS, Universit\'e Paris-Sud, UVSQ, Universit\'e Paris-Saclay,
91191 Gif-sur-Yvette, France}
\author{F. Qu\'er\'e}
\affiliation{LIDYL, CEA, CNRS, Universit\'e Paris-Saclay, CEA Saclay, 91 191 Gif-sur-Yvette, France\\}
\author{ H. Vincenti}
\email[]{henri.vincenti@cea.fr}
\affiliation{LIDYL, CEA, CNRS, Universit\'e Paris-Saclay, CEA Saclay, 91 191 Gif-sur-Yvette, France\\} 

\date{\today}

\begin{abstract}
Doppler harmonic generation of a high-power laser on a relativistic plasma mirror is a promising path to produce bright attosecond light bursts. Yet, a major challenge has been to find a way to generate isolated attosecond pulses, better suited to timed-resolved experiments, rather than trains of pulses. A promising technique is the attosecond lighthouse effect, which consists in imprinting different propagation directions to successive attosecond pulses of the train, and then spatially filtering one pulse in the far field. However, in the relativistic regime, plasma mirrors get curved by the radiation pressure of the incident laser and thus focus the generated harmonic beams. This increases the harmonic beam divergence and makes it difficult to separate the attosecond pulses angularly. In this article, we propose two novel techniques readily applicable in experiments to significantly reduce the divergence of Doppler harmonics, and achieve the generation of isolated attosecond pulses from the lighthouse effect without requiring few-cycle laser pulses. Their validity is demonstrated using state-of-the-art simulations, which show that isolated attosecond pulses with $10TW$ peak power in the X-UV range can be generated with PW-class lasers. These techniques can equally be applied to other generation mechanisms to alleviate the constraints on the duration on the laser pulses needed to generate isolated attosecond pulses. 
\end{abstract}

\maketitle

%%%%%%%%%%%%%%%%%%%%%%%%%%%%%%%%%%%%%%%%%%%%%%%%%%%%%%%%%%%%%%%%%%%%%%%%%%%%%%%%%%%%%%
%%%%%%%%%%%%%%%%%%%%%%%%%%%%%%%%%%%%%%%%%%%%%%%%%%%%%%%%%%%%%%%%%%%%%%%%%%%%%%%%%%%%%%

\section{Introduction}

High order harmonic generation of femtosecond lasers has been key to the advancement of attosecond science \cite{RevModPhysattoscience}. This physical process occurs when focusing an intense femtosecond laser in different media, such as atomic or molecular gases \cite{RevModPhysattoscience}, bulk crystals \cite{ghimire2011observation}, or overdense plasmas generated at the surface of solid targets \cite{RevModPhys.81.445, Thaury2007}. In all cases, the general picture is the same: due to the high laser intensity, the strong non-linear optical response of the medium to the incident field periodically distorts the waveform of the transmitted or reflected field, resulting in a spectrum of a high-order harmonics in the frequency domain. Filtering off the fundamental laser frequency, one can then obtain a train of sub-fs pulses in the time domain, provided the induced waveform distortion is localized in time within each laser optical cycle. 

%In gases, the physical process underpinning harmonic generation has been identified as the three-step electron recombination process in gas atoms \cite{corkum1993plasma} and occurs as long as the laser intensity is not too high $I<10^{14} W/cm^2$ (so that the gas is not fully ionized). 

In overdense plasmas, the harmonic generation processes require very high laser intensities ($I>10^{16} W/cm^2$) at which the initial solid target is turned into a so-called plasma mirror \cite{PhysRevE.69.026402,kapteyn1991prepulse} (abbreviated PM in the reminder of this article) that can specularly reflect the incident light. 
Two main harmonic generation processes on PM have been identified in the literature, depending on laser intensity. The first one called 'Coherent Wake Emission' (CWE) \cite{CWEORIGINAL} starts occurring at moderately high intensities ($I \simeq 10^{16} W/cm^2$) and is triggered by laser-driven electron bunches that excite collective electronic plasma oscillations in the density gradient between vacuum and the plasma bulk. 

The second mechanism, called 'Relativistic Oscillating Mirror' (ROM) \cite{Thaury2007,Dromey2006}, occurs at even higher intensities ($I>10^{18} W/cm^2$) at which the laser drives periodic oscillations of the PM surface at relativistic velocities. These periodic oscillations induce a Doppler effect on the reflected field, which is responsible for the waveform distortion mentioned above. As the ROM mechanism is not limited in terms of laser intensity, it is expected to produce very bright attosecond light pulses that could be used to perform attosecond pump - attosecond probe experiments on electron dynamics in matter. Yet, a major difficulty to overcome with this harmonic source is to produce isolated attosecond pulses, better suited to perform time-resolved experiments, rather than  trains of  pulses.

%at the laser frequency temporally distorts the reflected field by Doppler effect and generates high order harmonics of the incident laser in the reflected field \cite{rommodellichters,baeva11,Gonoskovmodel}. Filtering off the laser fundamental frequency, this high harmonic spectrum corresponds to a train of sub-fs pulses in the time domain.
Some evidence for the generation of such isolated attosecond pulses have recently been reported, using few-cycle long laser pulses to drive the laser-plasma interaction \cite{jahn2019towards,kormin2018spectral}. The short duration of the driving laser pulse, combined with the strong non-linearity of the generation process, ensures that one attosecond pulse only is produced if the highest harmonic orders are selected -a scheme called intensity gating. Such few-cycle laser pulses are however extremely difficult to produce at ultrahigh laser intensities: they require custom-made state-of-the-art laser systems, whose powers are still far below the present records achieved by more conventional systems delivering pulse durations of the order of $6$ to $15$ optical periods. Other approaches are thus needed to fully exploit the potential of relativistic plasma mirrors driven by multi-PW lasers.

More advanced gating techniques have been developed to generate isolated attosecond pulses, either to alleviate the constraint on the duration of the driving laser pulse, or to extend the range of harmonics that can be selected \cite{sansone2006isolated,goulielmakis2008single,feng2009generation,alhhenri,Wheeler2012, Heyl_2014,PhysRevLett.115.193903}. Among those, a general gating technique, called the attosecond lighthouse effect \cite{alhhenri,Wheeler2012,kimPhotonicStreakingAttosecond2013,quere2014applications}, consists in applying a controlled spatio-temporal coupling called Wavefront Rotation (WFR) \cite{Akturkreview} to the driving laser at focus: the direction of the incident laser light varies linearly in time along the femtosecond envelop of the laser pulse. Due to this temporal rotation, successive attosecond light pulses are emitted in slightly different directions. If the angular separation between two successive attosecond pulses is high enough, one can then obtain an isolated attosecond pulse in the far field by simply placing a slit to spatially select one pulse of the train. Considering that all attosecond pulses are emitted in the divergence cone of the incident laser, the total number of attosecond pulses that can be isolated with this scheme (or equivalently the maximum laser duration that can be used) is simply given by the ratio $\theta_L/\theta_n$ of laser and harmonic beam divergences. 

%For gas and CWE harmonics, this ratio is of the order of $3$ to $4$, allowing the generation of isolated attosecond pulses with a few-cycle laser pulse as demonstrated in \cite{kimPhotonicStreakingAttosecond2013,Wheeler2012}.

However, for Doppler harmonics produced by a ROM, it has been shown that laser radiation pressure curves the PM surface \cite{Dromey2009, Vincenti2014}, leading to an enhanced harmonic divergence and a ratio $\theta_L/\theta_n$ of the order of unity (for laser-plasma conditions optimizing harmonic generation). Obtaining isolated attosecond pulses through the lighthouse effect in the ROM regime would thus require high-power laser pulses with durations of at most two cycles, with limited benefits compared to conventional intensity gating. This has severely hindered the use of the attosecond lighthouse scheme in the ROM regime, for which no experimental demonstration has yet been reported. 

In this article, we propose two techniques to significantly reduce the divergence of Doppler harmonics, and implement the gating of isolated attosecond pulses with the attosecond lighthouse effect in the ROM regime. We emphasize that these schemes are not specific to the ROM mechanism -although they are particularly relevant in this case- but equally apply to any type of source of attosecond pulses. These two techniques require a simple tuning or tailoring of the driving laser phase or amplitude profile (on top of the applied WFR), which is in both cases achievable with current experimental know-how. This article is divided as follows: 

\begin{enumerate}[label=(\roman*)]
    \item In section {\rm II}, we  remind the limitations of the attosecond lighthouse effect in the ROM regime in a more quantitative way. 
    \item In section {\rm III}, we present a technique to reduce harmonic beam divergence by tuning the wavefront curvature of the incident laser.
    \item In section {\rm IV}, we present a second technique to reduce the harmonic beam divergence by tailoring the amplitude profile of the incident laser beam.
    \item In section {\rm V}, we perform a 3D numerical experiment with the Particle-In-Cell (PIC) code WARP+PXR to validate the first technique and provide quantitative estimates of the properties of the isolated attosecond pulses that could be obtained with a PW-class laser.
\end{enumerate}

\section{Limitations of the lighthouse effect in the relativistic regime}
\subsection{Separation criterion}

\begin{figure}[h]
\centering
\includegraphics[width=0.7\linewidth]{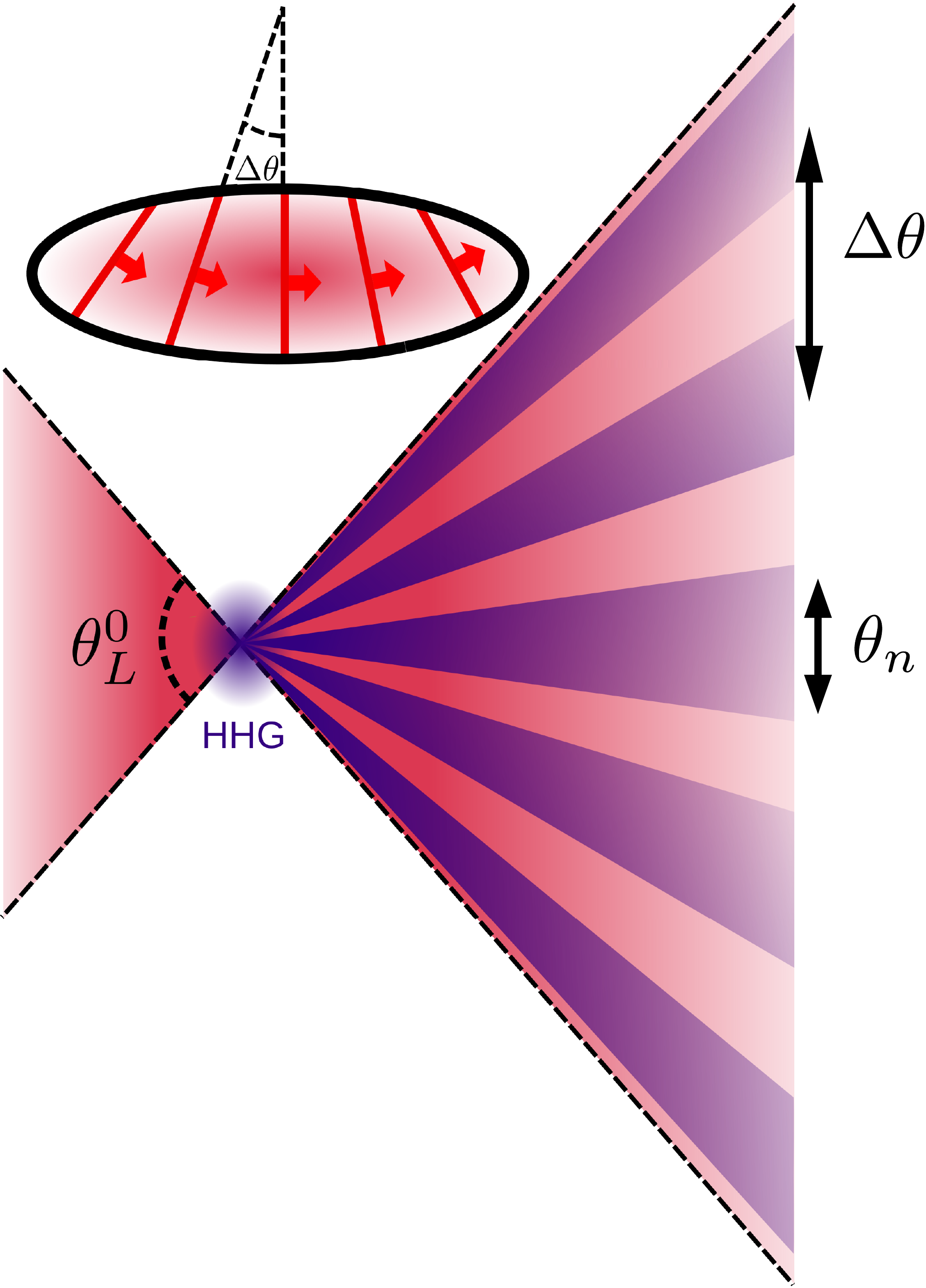}
\caption{Principle of the attosecond lighthouse effect: separation criterion. The incident laser beam is shown in red and the multiple attosecond beamlets generated through the highly non-linear interaction are shown in purple. The wavefront rotation (WFR) of the laser field at focus is sketched in the upper inset.}
\label{lighthouseprinciple}
\end{figure}

The spatial separation of two successive attosecond pulses is possible provided that the angular offset $\Delta \theta$ between two successive attosecond pulses, induced by WFR, is larger than the divergence $\theta_n$ of the harmonic beam [cf. Fig. \ref{lighthouseprinciple}]: 
\begin{equation}
    \Delta\theta = v_r\Delta T \geqslant \theta_n
\end{equation}
where $v_r$ is the WFR velocity of laser wavefronts at focus and $\Delta T$ is the time delay between the emission of two successive attosecond pulses. $\Delta T$ is equal to the laser period $T_0$ for attosecond pulses emitted on relativistic PMs at oblique incidence. The velocity $v_r$ can be estimated by stating that during the entire laser pulse, the laser wavefront rotate by an angle $\theta_0$ corresponding to the divergence of the focused laser beam. This leads to $v_r= \theta_0/N T_0 $, where $N$ is the number of optical cycles in the incident laser pulse. The separation criterion then writes:  
\begin{equation}
     \theta_n \leqslant \theta_0/N
\end{equation}
For a given harmonic beam divergence, the above criterion gives the maximum number of optical cycles $N_{max}\approx \theta_0/\theta_n$ of the incident laser pulse up to which it is possible to induce a clear angular separation of successive attosecond pulses in the far field with the lighthouse effect.   

\subsection{Current limitations in the relativistic regime}
\label{currentlimsub}

\begin{figure}[h]
\centering
\includegraphics[width=\linewidth]{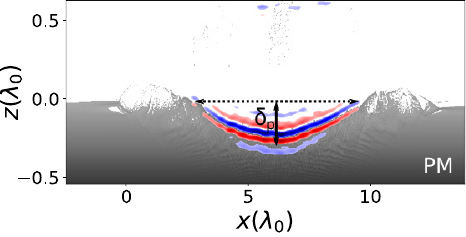}
\caption{PM denting by laser radiation pressure. This figure shows simulation results of a 2D Particle-In-Cell (PIC) simulation. In this simulation, the laser impinges on the PM at a $45^\circ$ angle of incidence. The PM has an exponential density profile of gradient scale length $L=\lambda_0/15$. The simulation results are displayed in a special Lorenz Boosted frame \cite{bourdierpaper} where the laser is at normal incidence on the PM, thereby simplifying the visualization of simulation results. The gray color scale represents a snapshot of the PM electron density. The color scale represents the generated Doppler harmonic field (zoom on a single attosecond pulse within the generated train of pulses), where we filtered harmonic orders $8$ to $22$.}
\label{dentingbourdier}
\end{figure}

At very high laser intensities, the spatially-inhomogeneous radiation pressure exerted by the incident laser field induces a denting of the PM surface, resulting in a curvature of this surface \cite{wilks1992absorption,Vincenti2014}. This in turn results in a curvature of the harmonic wavefronts, leading to a tight focusing of the harmonic beam in front of the PM surface \cite{vincenti2019achieving}.  This is illustrated on Fig. \ref{dentingbourdier} showing a snapshot of the PM electron density (gray scale) and of the reflected field (color scale) frequency-filtered from harmonic orders 8 to 22, obtained from PIC simulations with the pseudo-spectral 3D PIC code WARP+PXR \cite{vay2012novel,vay2013domain,vincenti2016detailed, blaclard2017pseudospectral, vincenti2018ultrahigh,vincenti2017efficient,vincenti201717,kallalahps}.

 This significantly increases the divergence of the harmonic beam. Assuming a Gaussian laser and harmonic beams, it can be shown that the divergence $\theta_n$ of the $n^{th}$ harmonic beam writes \cite{Vincenti2014}: 
\begin{equation}
\theta_n = \theta_n^{0} \sqrt{1+(n\psi)^2}
\end{equation}
where $\theta_n^{0}=\lambda_n/\pi w_n$ is the diffraction-limited divergence (i.e. without focusing by the PM) and $\psi$ is defined as:  
\begin{equation}
\psi = \frac{2 \pi} {\cos\theta}\left(\frac{w_n}{w_0}\right)^2\frac{\delta_p}{\lambda_0}
\end{equation}
with $w_n$ the harmonic beam waist on the PM, $w_0$ the laser waist, $\theta$ laser angle of incidence on the PM, $\lambda_0$ the laser wavelength and $\delta_p$ the parameter describing the denting of the curved PM as defined on Fig. \ref{dentingbourdier}. For a high-enough laser amplitude $a_0\gg 1$, this last parameter is given by: 
\begin{equation}
\delta_p \approx 4L \cos\theta^2
\label{deltap}
\end{equation}
 where $L$ is the scale length of the density gradient at the PM-vacuum interface. This scale length is a key parameter of the interaction and is much shorter than the laser wavelength $\lambda_0$ in the regime of efficient harmonic generation. As $L$ increases, the local plasma density of the PM decreases, which makes it easier for the incident laser to dent the PM surface by radiation pressure. This results in an increased harmonic beam divergence for large $L$. Assuming $w_n=w_0$, one can show that for harmonic orders that are focused by the PM (i.e. such that $n\psi\gg 1$):
\begin{equation}
\frac{\theta_L}{\theta_n} = \frac{\lambda_0}{8\pi\cos\theta L}
\end{equation}
For a gradient scale length in the range $L\approx \lambda_0/20-\lambda_0/8$ that optimizes harmonic generation efficiency for laser angles of incidence between $60^\circ$ and $45^\circ$ \cite{PhysRevLett.110.175001,chopinaeauprx}, $N_{max}=\theta_0/\theta_n$  is thus of the order of $1$. This shows that generating isolated attosecond pulses with the lighthouse effect in laser-plasma conditions that are optimal for Doppler harmonic generation is very challenging, as it would require laser pulses with a duration of the order of one optical cycle. Such single-cycle pulses are extremely hard to obtain for high-power lasers, which usually rather provide pulses with durations between $15$ and $40$ fs (i.e. from $6$ to $15$ laser periods for a central wavelength $\simeq 800$ nm). 

To break this barrier, we hereby propose two techniques to significantly reduce the harmonic beam divergence $\theta_n$, by combining WFR with an additional shaping of the spatial phase or amplitude profile of the incident laser beam.

% Brief reminder on the attosecond lighthouse effect and the separation criterion
% Focusing by radiation pressure : detail 
% Give orders of magnitude of Delta theta/theta_n: no angular separation of two successive pulses unless for a few cycle laser
% Having a few cycle laser in the 100TW and up to the PW regime is currently not feasible
% In this paper, we propose two new techniques to mitigate the radiation pressure focusing and reduce harmonic beam divergence. 

\section{Reduction of harmonic divergence by tuning the curvature of laser wavefronts}

\subsection{General principle}

\begin{figure}[h]
\centering
\includegraphics[width=\linewidth]{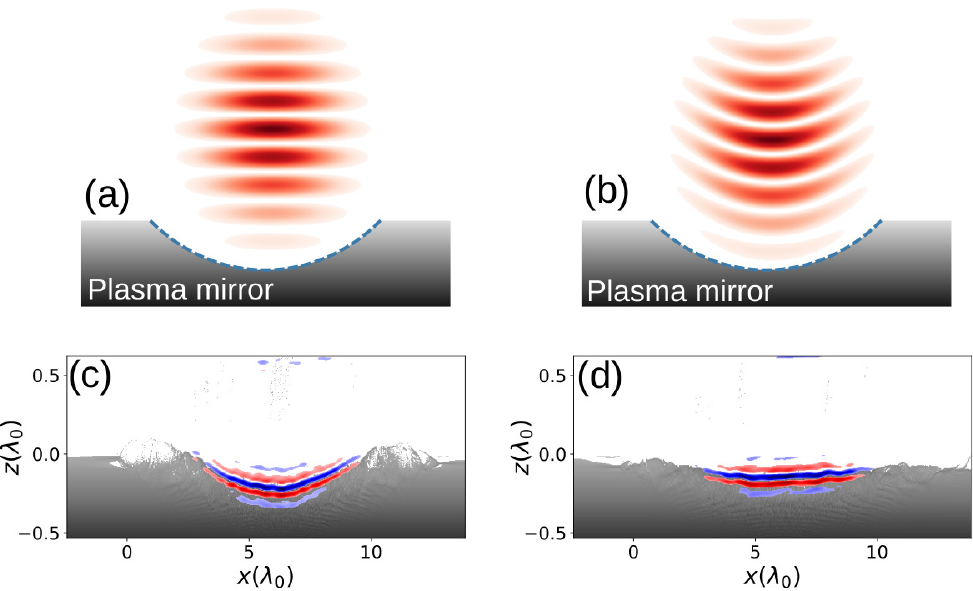}
\caption{ 2D PIC simulations of attosecond pulse emission by a relativistic PM. Simulation results are displayed in the same boosted frame as in Fig. \ref{dentingbourdier}, where the laser is normally incident on the PM. Panels (a) and (b): schematic representation of driving laser wavefronts. Panels (c) and (d): The color scale represents the electromagnetic field of one attosecond pulse of the train, when the PM surface is placed at the best focus of the laser beam, and at a distance $\delta z=0.6 Z_r$ after this best focus ($Z_r$ Rayleigh length of the laser beam), respectively. The gray color scale represents the electron density of the PM. 
}
\label{defocbourdier}
\end{figure}

The general principle of the technique is sketched on Fig. \ref{defocbourdier} and consists in placing the PM slightly away from the laser best focus, so that the incident laser wavefronts are curved and compensate for the wavefront curvature induced by the PM on the reflected field. To achieve this, the PM must be placed at a defocusing distance $\Delta z$ such that the incident laser wavefronts are diverging in the PM plane, as illustrated on Fig. \ref{defocbourdier} (b).  This compensation scheme has been demonstrated experimentally in \cite{Vincenti2014}, in the absence of WFR.

 In the following, we develop a theoretical model from which we can predict the optimal defocusing distance that maximizes the ratio between the angular separation of successive attosecond pulses and the harmonic beam divergence, as a function of laser-plasma parameters. 

\subsection{Model for a Gaussian laser beam with WFR out of focus}
\label{model_field_wfr}

Determining the angular separation of attosecond pulses with the defocusing technique first requires to know the velocity of WFR out of focus. In this section, we derive the analytical expression of WFR at a distance $\Delta z$ from the laser best focus (located at $\Delta z=0$). 

We first remind the properties of a beam with WFR at best focus, considering a Gaussian laser beam in space and time. WFR is a spatio-temporal coupling, whose spatio-spectral counterpart is Spatial Chirp (SC), i.e. the focusing of different frequencies within the pulse bandwidth at different transverse positions. These couplings at focus can be easily induced by applying a spatio-temporal coupling known a Pulse Front Tilt (PFT)  \cite{alhhenri,Akturkreview} on the collimated beam before focusing. PFT corresponds to a tilt between the wavefront and energy front of the beam, and is quantified by a parameter $\xi$, typically expressed in $fs/mm$, which can for instance be controlled by rotating one of the gratings in the compressor of a Chirped Pulse Amplification laser.   

The laser propagates along the direction $z$ and the SC is induced along the transverse direction $x$. The laser electric field at focus writes: 
\begin{multline}
E(x,t,z=0) = E_0 e^{-x^2/ w^2_{\xi}} e^{-t^2/\tau_{\xi}^2} e^{i \omega_0 t +i k_0 v_\xi x t}
\label{Extz0}
\end{multline}
where $E_0$ is the field amplitude and $v_\xi$ is the rotation velocity of the wavefront at the laser best focus, given by: 
\begin{equation}
    v_\xi = \frac{\xi/\xi_0}{1+(\xi/\xi_0)^2}v_{0}
\end{equation}
with : 
\begin{equation}
    v_{0} =\frac{2}{\tau_0k_0w_0}
\end{equation}
the maximum WFR rotation velocity reached for an optimal PFT $\xi=\xi_0$ \cite{alhhenri} given by:
\begin{equation*}
    \xi_0 = \frac{\tau_0}{w_i}
\end{equation*}
where $w_i$ is the laser waist before focusing (i.e. the size of the unfocused beam), and $\tau_0$ is the local duration of the unfocused beam \footnote{The local duration is the duration of the pulse at one transverse position in the beam. It should be distinguished from the global duration, which is the duration of the pulse when energy is spatially integrated all across the beam profile \cite{Bourassin-Bouchet:11}}. At best focus, the local pulse duration is no longer $\tau_0$, but is increased to: 
\begin{equation}
    \tau_\xi = \tau_0 \sqrt{1+(\xi/\xi_0)^2}
\end{equation}

In addition, due to SC, the spectrally-integrated laser focal spot is elongated along $x$. The laser waist at focus along the direction of SC is thus: \begin{equation*}
    w_\xi = w_0\sqrt{1+(\xi/\xi_0)^2}
\end{equation*}
while it remains equal to $w_0$ in the other transverse direction. The corresponding elliptical shape of the focal spot with WFR will be shown to have important consequences on the spatial properties of the generated attosecond pulses.

To get the expression of WFR at an arbitrary longitudinal position $\Delta z$, we propagate the field (initially known at $\Delta z=0$, see Eq. \ref{Extz0}) at position $\Delta z$ using a plane wave decomposition, and obtain (see derivation in appendix):
\begin{multline}
\label{extdz}
E(x,t'=t-\Delta z/c,z=\Delta z)   \\  \propto e^{- \dfrac{k_0 x^2}{2 \big (Z_\xi+i\Delta z\big)}} \times e^{-\dfrac{t'^2}{\tau^2_{\xi}}  \left[1+\dfrac{(\xi/\xi_0)^2 \Delta z^2}{\Delta z^2+Z_\xi^2}\right]} \\ 
 \times e^{ -\dfrac{i(\xi/\xi_0)^2t'^2}{\tau^2_\xi\left(Z_\xi/\Delta z+\Delta z/Z_\xi\right)}} \times e^{i\left[\omega_0 + \dfrac{\zeta x}{1+i \dfrac{ \Delta z}{Z_\xi } }\right]t'}
\end{multline}
where $Z_{\xi}=\pi w_\xi^2/\lambda_0$ is the laser Rayleigh range in the plane of WFR and $\zeta= k_0 v_\xi $ is the SC of the laser beam at focus defined in the spatio-spectral domain. The physical meaning of the different terms in Eq. \ref{extdz} is explained in the appendix. 

The WFR effect is all contained in the phase of the last exponential. It can be shown (see appendix) that the WFR velocity $v_\xi(\Delta z)$ at a distance $\Delta z$ from focus is given by:
\begin{equation}
    v_\xi(\Delta z) = \dfrac{v_\xi}{1+\left(\frac{\Delta z}{Z_\xi}\right)^2}
    \label{vxi}
\end{equation} 
The above equation shows that the WFR velocity decreases with $\Delta z$. However, as long as $\Delta z$ is smaller than the laser Rayleigh range $Z_\xi$, this decrease is limited. The above formula can be used to derive the angular separation of attosecond pulses generated from a target placed at distance $\Delta z$ from best focus, through:  
\begin{equation}
\Delta \theta_\xi (\Delta z) = v_\xi(\Delta z) T_0
\label{eqdtheta}
\end{equation}
In the following, we combine this result with a model of harmonic divergence $\theta_n(\xi,\Delta z)$ to find the optimal defocusing distance beyond which $\Delta \theta_\xi (\Delta z)>\theta_n(\xi,\Delta z)$. 

\subsection{Model for the harmonic beam spatial phase and divergence}

The total harmonic phase in the PM plane along the direction of WFR can be written \cite{Vincenti2014} : 
\begin{equation}
\begin{split}
\phi_n(x)&=\frac{2\pi n }{\lambda_0}\left[ \frac{x^2}{2R_\xi(\Delta z)}+\frac{x^2}{2 f_{\xi}(\Delta z)\cos\theta}\right] \\
\end{split}
\label{tothhgphase}
\end{equation}
%where $x$ is a coordinate transverse to the direction of propagation of the harmonic field (i.e. the specular reflection direction).
The left term at the right hand side of Eq. \ref{tothhgphase} accounts for the spatial curvature of the laser wavefront on the PM. If $\Delta z\rightarrow 0$,  $|R_\xi|\rightarrow \infty$ and this term vanishes -this corresponds to the standard case where PM is placed at best focus. In this case, the harmonic phase is only governed by the second term on the right hand side, corresponding to the phase term induced by the PM curvature, where: 
\begin{equation}
    f_\xi(\Delta z) = \frac{w_\xi(\Delta z)^2}{4L\cos\theta^2}
\end{equation}
is the focal length of the curved PM as derived in \cite{Vincenti2014}. 

When $\Delta z<0$ (i.e. the laser is focused before the PM), Eq. \ref{tothhgphase} shows that the mitigation of the phase term induced by the PM curvature arises from two physical effects: 
\begin{enumerate}[label = (\roman*)]
    \item the negative quadratic phase term associated to the wavefront curvature of the laser beam, which tends to compensate the opposite curvature induced by the PM surface. 
    \item an increase of the PM focal length $f_\xi$, resulting from the increase of $w_\xi(\Delta z)$ as the target surface is moved away from best focus. 
\end{enumerate}

Both effects lead to a reduction of the harmonic beam divergence. Using Eq. \ref{tothhgphase} and following the same approach as in \cite{Vincenti2014}, one can derive a modified model of the harmonic beam divergence for a defocusing distance $\Delta z$: 
\begin{multline}
    \theta_n(\Delta z,\xi) = \theta_n^0(\Delta z,\xi)  \\
    \times \sqrt{1+\left[n\psi_\xi(\Delta z)\right]^2\left[1+\frac{f_\xi(\Delta z)\cos\theta}{R_\xi(\Delta z)}\right]}
    \label{divhhgdz}
\end{multline}
with : 
\begin{equation}
     \psi_\xi(\Delta z)   = \frac{2\pi}{\cos\theta}\left[\frac{w_n(\Delta z,\xi)}{w_\xi(\Delta z)}\right]^2\frac{\delta_p}{\lambda_0}
\end{equation}
and $\delta_p$ defined by Eq. \ref{deltap}. $\theta_n^0(\Delta z,\xi)=\lambda_n/\pi w_\xi(\Delta z)$ corresponds to the diffraction-limited divergence of harmonic beams when those are generated with the target at a distance $\Delta z$ from best focus. In the above Eq. \ref{divhhgdz}, the only quantity for which there is currently no analytical model is the harmonic source size $w_n$, which depends on laser-plasma parameters. In the limit of ultra-high laser amplitudes $a_0\gg 1$ in the plane of generation, one can reasonably assume $w_n(\xi,\Delta z)\approx w_\xi(\Delta z)$ for a broad range of harmonic orders \cite{vincenti2019achieving}. At lower intensities, one has to rely on PIC simulations for a more accurate estimation of this quantity. 

\subsection{Optimal defocusing distance}

The optimal defocusing distance $\Delta z_\xi$ is reached when the ratio $\eta_\xi(\Delta z) = \Delta\theta_\xi(\Delta z)/\theta_n(\xi,\Delta z)$ is maximized, where $\Delta\theta_\xi$ is given by Eq. \ref{eqdtheta} and $\theta_n$ by Eq. \ref{divhhgdz}. For a fixed value of PFT $\xi$, $\Delta z_\xi$ can simply be found by solving the following equation : 
\begin{equation}
    \dfrac{d \eta}{d\Delta z}(\Delta z_\xi) = 0
    \label{eqdeltaxi}
\end{equation}

\subsubsection{Optimal defocusing distance for a fixed PFT}

\begin{figure}[h]
\centering
\includegraphics[width=\linewidth]{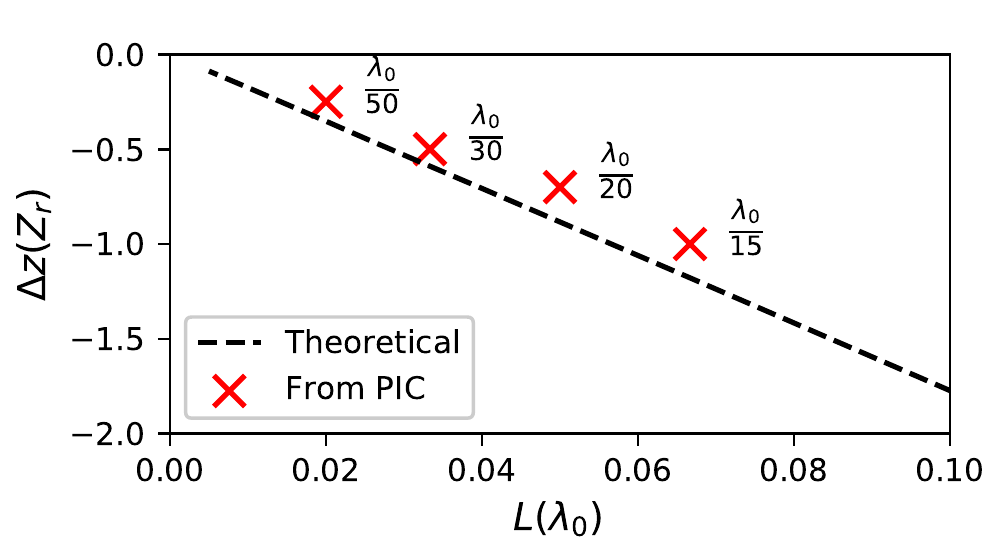}
\caption{Optimal defocusing distance as a function of the gradient density scale length $L$ at the PM surface. The red crosses are obtained from 2D PIC simulations, and the black dashed line corresponds to the prediction of Eq.\ref{analyticalmodelDeltaz} for a laser amplitude $a_0=30$ (without WFR), an angle of incidence $\theta=45^\circ$, a laser duration $\tau_0 = 16fs$ and a PFT parameter $\xi=\xi_0$.}
\label{moddelvalidation}
\end{figure}

Eq. \ref{eqdeltaxi} is a third-order polynomial equation in $\Delta z$ that can be exactly solved numerically. This optimal distance $\Delta z_\xi$ depends on laser and plasma parameters -mostly the laser angle of incidence $\theta$ and the plasma density gradient scale length $L$. 

Actually, one can find a very good analytical approximation of $\Delta z_\xi$ by using the facts that 1- $\Delta\theta_\xi$ hardly varies with $\Delta z$ as long as $\Delta z\ll Z_\xi$,  and 2- $\eta_\xi$ is maximum when $\theta_n$ is minimum (i.e. close to its diffraction-limited value $\theta_n^0$). The second condition occurs when the total harmonic phase $\phi_n(x)$ defined in Eq. \ref{tothhgphase} is constant,  \textit{i.e.} when: 
\begin{equation}
    R(\Delta z_\xi) = -f_p(\Delta z_\xi)\cos\theta
\end{equation}
Note that according to Eq.\ref{divhhgdz}, this condition is indeed the one minimizing $\theta_n(\Delta z,\xi)$.
Solving the above equation yields: 
\begin{equation}
    \Delta z_\xi = -Z_\xi\left[\dfrac{1}{2 \Lambda_p}+\sqrt{\dfrac{1}{4\Lambda_p^2}-1}\right]
    \label{analyticalmodelDeltaz}
\end{equation}
with $\Lambda_p = 4\pi \cos\theta L/\lambda_0$ a dimensionless parameter that only depends on the laser angle of incidence and PM gradient scale length $L$. This solution exists as long as $\Lambda_p\leqslant 1/2$. Beyond this limit value, the PM denting phase is not fully compensated by the incident laser phase and one has to rely on the numerical resolution of Eq. \ref{eqdeltaxi} to determine $\Delta z_\xi$. 

To check the validity of this model, we ran a parameter scan of 2D Particle-In-Cell (PIC) simulations with the WARP+PXR code where we varied the gradient scale length $L$ and the PM defocusing distance $\Delta z$. For each gradient scale length $L$, we extracted from these simulations the optimal defocusing distance $\Delta z_\xi(L)$ that maximizes the ratio $\eta_\xi$. These are indicated as red markers on Fig. \ref{moddelvalidation}. As expected, when $L$ is increased, the required defocusing distance increases, due to the augmentation of PM denting (and thus harmonic divergence) as shown in Eq. \ref{deltap}. One can see on Fig. \ref{moddelvalidation} that optimal defocusing distances obtained from simulations match those predicted by the analytical model (black dashed line) given by equation Eq. \ref{analyticalmodelDeltaz}, except for a small overall offset.

\begin{figure}[h]
\centering
\includegraphics[width=1\linewidth]{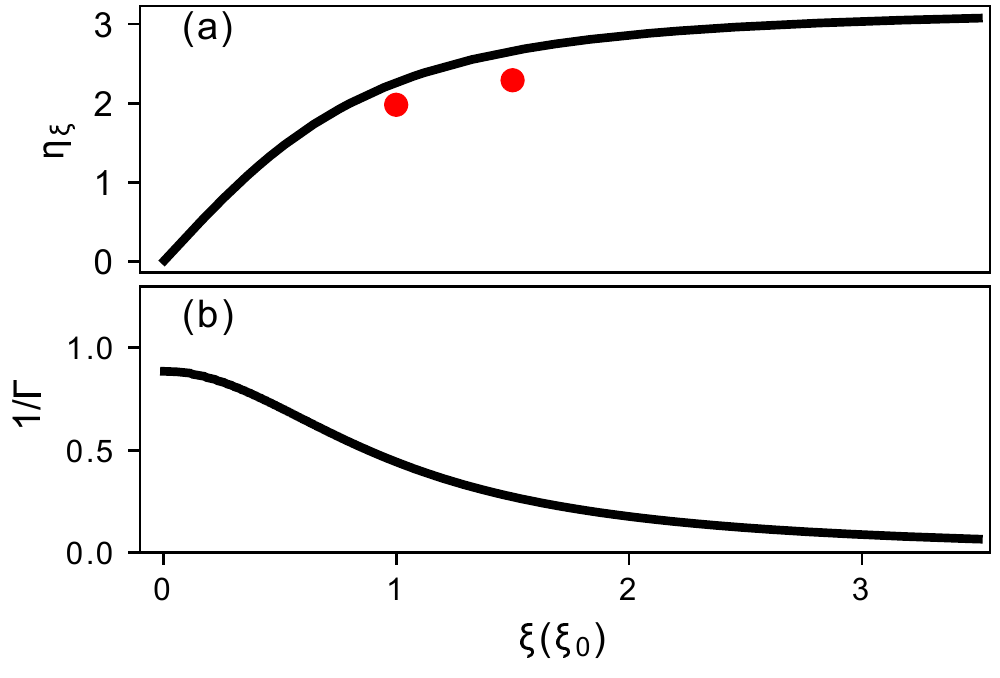}
\caption{ Panel (a): Evolution of $\eta_\xi$  as a function of $\xi$. The black curve represents $\eta_\xi$ as given by Eq. \ref{etaxi_model} for harmonic order $n=20$. Red circles correspond to $\eta_{\xi}$ obtained from 2D PIC simulations for harmonic order $n=20$. In both cases, we assumed an angle of incidence $\theta_0=55^\circ$, a laser duration $\tau_0 = 16fs$ (without WFR), a PM gradient scale length $L=\frac{\lambda_0}{20}$ and a normalized laser amplitude $a_0=30$ (at best focus, without WFR). Panel (b): Evolution of the actual laser amplitude at the target surface obtained using Eq. \ref{intensity_loss}, with WFR and optimal defocusing.}
\label{etaxi}
\end{figure}

The main price to pay in experiments for this optimisation of the ratio $\eta_\xi(\Delta z)$ is a reduction of the peak intensity on target, and hence of the harmonic generation efficiency. This intensity reduction can be deduced from the beam waist and pulse duration provided by Eq. \ref{extdz}, and writes for the optimal defocusing distance $\Delta z_\xi$:
\begin{equation}
    \Gamma = \dfrac{1}{1+(\xi/\xi_0)^2}\times \dfrac{1}{{1+\left[\frac{1}{2 \Lambda_p}+\sqrt{\frac{1}{4\Lambda_p^2}-1}\right]^2}}
    \label{intensity_loss}
\end{equation}
The first term on the right hand side of the above equation is the intensity decrease due to the introduction of PFT, while the second term is the intensity decrease due to the laser defocusing. For $\xi=\xi_0$ (maximizing WFR at laser focus) and realistic laser-plasma parameters optimizing harmonic generation ($\theta=55^o$, $L\approx\lambda_0/15$) \cite{chopinaeauprx} one finds $\Gamma\approx 3$.  This intensity reduction does not compromise the use of this scheme for high-order harmonic generation from relativistic plasma mirrors:  indeed, the current generation of high-power femtosecond lasers can now deliver intensities $I>10^{20}W/cm^2$, more than two orders of magnitude higher than the threshold for Doppler harmonic generation ($I\simeq 10^{18}W/cm^2$).

\subsubsection{Effect of PFT on the angular separation of attosecond pulses}

\label{zetabest}

\begin{figure}[h]
\centering
\includegraphics[width=0.95\linewidth]{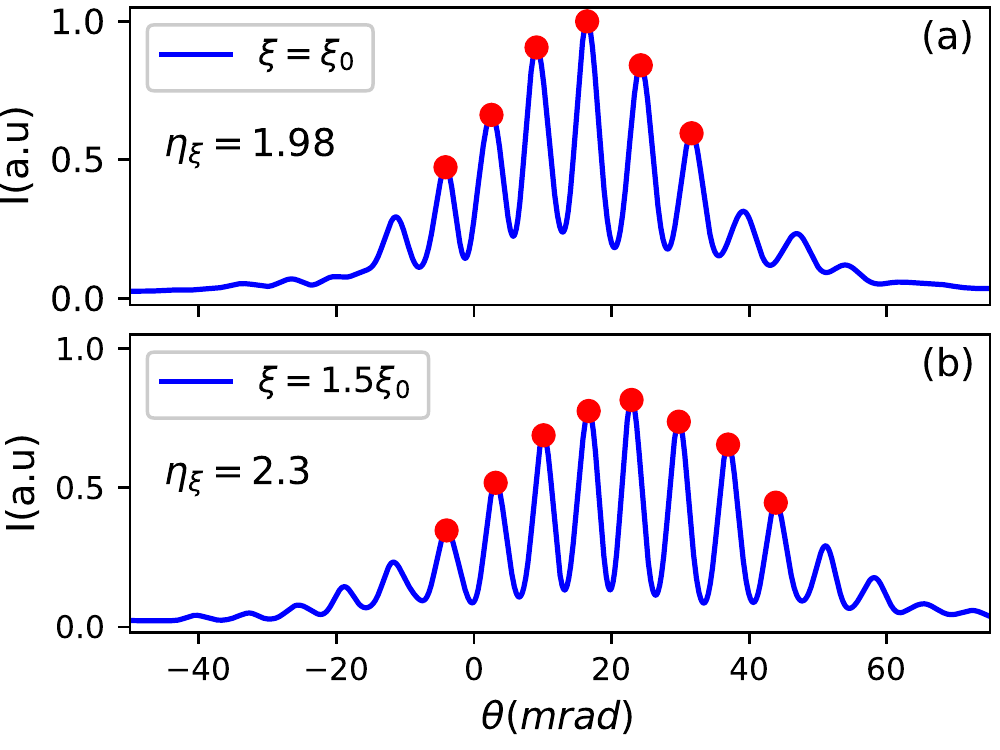}
\caption{ Angular profiles of Doppler harmonic beams (orders $20$ to $30$) obtained from 2D PIC simulations for two different pulse-front tilt values: $\xi =\xi_0$ (panel (a)) and $\xi =1.5\xi_0$ (panel (b)). All other laser and plasma parameters are kept constant otherwise: $\theta_0 = 55^\circ$, $L= \frac{\lambda_0}{20}$, a laser duration $\tau_0 = 16fs$ (without WFR, at best focus) and $a_0=30$ (at best focus, without WFR). On each panel, the red dots highlight the direction of emission of the successive attosecond pulses of the train. }
\label{compare_xis}
\end{figure}

At the optimal defocusing distance $\Delta z = \Delta z_\xi$, harmonic divergence reaches its diffraction-limited value and the ratio $\eta_\xi$ (obtained using results of Eq. \ref{eqdtheta} and \ref{analyticalmodelDeltaz}) writes : 
\begin{equation}
    \eta_\xi =  \frac{2\xi/\xi_0}{\sqrt{1+(\xi/\xi_0)^2}}\frac{v_0T_0/\theta_n^0}{\sqrt{1+\left[\frac{1}{2\Lambda_p}+\sqrt{\frac{1}{4\Lambda_p^2}-1}\right]^2}}
    \label{etaxi_model}
\end{equation}

The evolution of $\eta_\xi$ with $\xi$ as given by Eq. \ref{etaxi_model}  is displayed in Fig. \ref{etaxi} (a). For $\xi< \xi_0$, when $\xi$ increases, the WFR velocity increases and the diffraction-limited divergence $\theta_n$ decreases (due to the increase of the laser waist), resulting in a rapid growth of the ratio $\eta_\xi$ when $\xi$ increases. For $\xi> \xi_0$, the rotation velocity now slowly decreases with $\xi$ but the diffraction-limited divergence $\theta^0_n$ keeps decreasing. This still leads to a net increase of $\eta_\xi$ when $\xi$ is increased, yet at a slower pace.

Attosecond pulses are angularly separated when $\Delta \theta_\xi>\theta_n$, i.e. $\eta_\xi>1$. This occurs starting at $\xi\approx 0.3\xi_0$. Further increasing $\xi$ leads to a better angular separation quality of attosecond pulses as illustrated on panels (a) and (b) of Fig. \ref{compare_xis},  at the cost of a further reduction of the laser intensity on target (see Fig.\ref{etaxi} (b)). The amount of PFT $\xi$ needed ultimately depends on the contrast ratio desired between the main spatially filtered attosecond pulse and the portions of satellite attosecond pulses that angularly-overlap with the main pulse.

\section{Shaping the laser beam spatial intensity profile}
In this section, we propose a second technique to reduce the divergence of Doppler harmonics, and achieve a good angular separation of attosecond pulses via the attosecond lightouse effect. Compared to the defocusing technique, the main advantage of this second scheme is that the target surface is kept at the laser best focus, where the beam intensity profile is usually of much better quality than out of focus and the WFR velocity is optimal. Yet, as we show in this section, this technique comes with an additional experimental complexity. 
\subsection{General principle}

\begin{figure}[h]
 \centering
\includegraphics[width=0.6\linewidth]{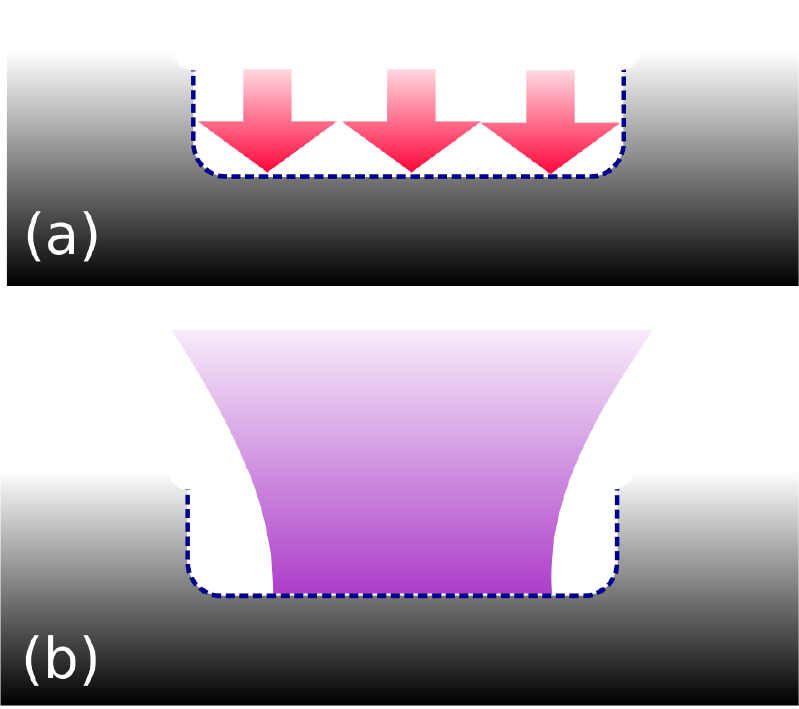}
\caption{  Illustration of (a) PM denting for a spatially-flat laser beam intensity profile and (b) of the Doppler harmonic beam generated by such a flat PM. }
\label{schemnovel}
\end{figure}

The general principle of this second technique, illustrated on Fig. \ref{schemnovel}, relies on shaping the spatial intensity profile of the incident laser beam at focus.  By flattening this profile (cf. panel (a)), one can suppress, or at least reduce, the laser-induced PM curvature, and thus mitigate the associated increase of  harmonic divergence (cf. panel (b)). 

\begin{figure}[ht]
 \centering
  \includegraphics[width=\linewidth]{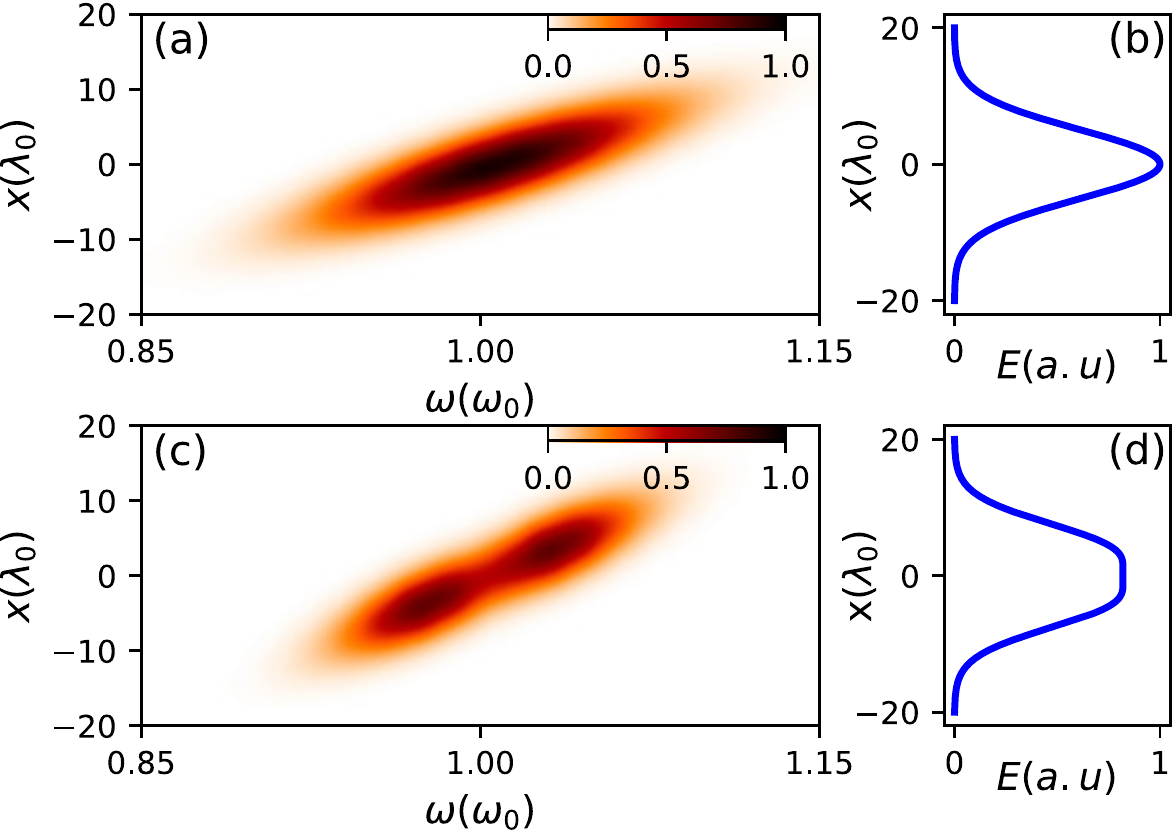}
 \caption{ Panel (a): Intensity of a laser beam with SC (but without any spectral shaping) in the spatio-spectral domain $(x,\omega)$ Panel (b): Spectrally-integrated spatial profile of this beam along the $x$ direction.  Panel (c) and (d): Same quantities, now with the application of the spectral shaping of Fig. \ref{newfig1D}. }
 \label{metrs}
\end{figure}

%\begin{figure}[h]
%\centering
%\includegraphics[width=\linewidth]{figures/merged_fr_all_spaces.eps}
%\includegraphics[width=\linewidth]{contrast_xw.eps}
%\caption{ Panel (a): Intensity of a spatially chirped laser beam at focus, in the spatio-spectral space $(x,\omega)$. Panel(b): Local spectrum of the spatially-chirped laser beam at different transverse positions $x$.} %Panel(c): The same field in the spatio-temporal space $(x,t)$ with (panel(b)) three slices of the field at different transverse positions.}
%\label{mergedallspaceswfr}
%\end{figure}

\begin{figure}[h]
 \centering
\includegraphics[width=\linewidth]{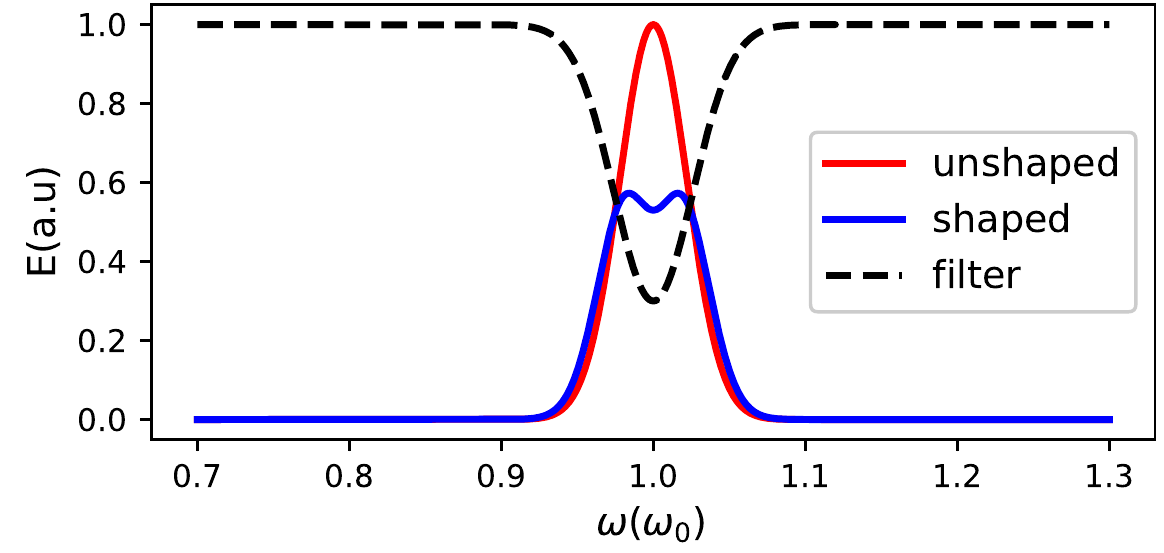}
\caption{Effect of spectral shaping on the laser spectrum. The red curve is the laser spectrum without any spectral shaping. Using a programmable acousto-optic filter, the gain modulation function shown in black is applied to the whole beam, leading to the spectrum plotted in blue. This beam is then shaped spatio-temporally (application of PFT) and focused to induce the attosecond lighthouse effect on target. }
\label{newfig1D}
\end{figure}

%\begin{figure*}[ht]
% \centering
  %\includegraphics[width=0.9\linewidth]{merged_process_flat.eps}
 %\caption{ Panel (a): Laser profile in spatio-spectral domain $(x,\omega)$ after filtering. Panel (b): local laser spectrum at three different positions across focus $x=0$ (green curve), $x = 0.75 w_{\xi}$ (blue curve) and $x = -0.75w_{\xi}$ (red curve). The black dashed line represents the filter function applied in the spectral domain. Panel (c): spatio-temporal profile of laser beam after spectral filtering. Panel (d): the red curve represents the laser spatial profile at $t=0$. The blue curve represents the same profile without any frequency filtering. Comment: il faut faire attention avec cette figure. Ce qui est tricky, c'est qu'on shape le spectre du laser de façon homogène sur tout le faisceau, avant focalisation, quand le spectre ne dépend pas de la position dans le faisceau. Mais une fois focalisé, le spectre dépend de la position. Sur le panel c, on mélange 2 choses, et ça peut être confusing : les spectres locaux au foyer, et la fonction filtre, qui elle est appliquée uniformément à tout le faisceau. Je ne pense pasque ce soit une bonne idée. A quoi sert le panel b au final ? Je suggère de revoir cette figure, en la faisant plus simple. Une comparaison des  paneaux a et d seulement, avec et sans filtrages serait plus parlante. A discuter. }
% \label{metrs}
%\end{figure*}

Such a top-hat spatial profile of the laser intensity at focus can in principle be obtained by tailoring the spatial phase of the beam before focusing, using either a simple phase plate \cite{boutu2011high} or a set of optical paths mirrors \cite{dubrouil2011controlling}. These schemes have however proved difficult to  implement efficiently on  high-power femtosecond lasers, even in the TW range. Moreover, they no longer apply in the conditions of the attosecond lighthouse gating scheme: WFR is unavoidably associated to SC, and the laser spatial intensity profile at focus is then partly determined by the laser spectrum. Pure spatial shaping techniques are then no longer suitable to control the spatial intensity profile at focus.

In the following, we present a technique that takes advantage of this coupling between spatial and spectral degrees of freedom in the attosecond lighthouse scheme. This could be used in a rather straightforward way in experiments to flatten the spatial intensity profile of a laser pulse with WFR, and relax the constraints for the generation  of isolated attosecond pulses with the lighthouse scheme. 

\subsection{Flattening the spatial intensity profile of a laser pulse with WFR}

As explained in section \ref{model_field_wfr}, WFR in the space-time domain corresponds to SC in the space-frequency domain [cf. Fig. \ref{metrs} (a)]. In the presence of SC, the laser central frequency $\omega_0$ varies as a function of the transverse position at focus $x$. This implies that in the limit of strong SC along direction $x$, the laser spatial profile along $x$ at focus actually corresponds to the spectral profile of the laser pulse -just as in the focal plane of a spectrometer.

As a result, one could exploit SC to tailor the spatial intensity profile of the laser beam at focus, simply by shaping the laser spectrum. % the relative amplitude of the frequency components of the initial laser pulse (before focusing), one could thus exploit SC at focus to flatten the spatial intensity profile of the laser beam at focus. 
More precisely, this profile could be flattened by damping the central frequency of the laser pulse as illustrated on Fig. \ref{newfig1D}. Such a spectral shaping is nowadays possible and rather straightforward using programmable acousto-optic modulators -such as the Dazzler \cite{dazzler1,dazzler2}- placed in the front end of high-power laser systems.

In order to simulate this scheme, we used the following frequency filter to damp the central laser frequency:

\begin{equation}
\begin{split}
 G(\omega) = 1 - \alpha e^{ - \big[ (\omega-\omega_0)\times \frac{\tau_0}{\beta} \big]^2 }
\label{eqfilterdamp}
\end{split}
\end{equation}
$\alpha$ and $\beta$ are tuning parameters that are used to control the amplitude as well as the standard deviation of the filter gain function. Fig. \ref{newfig1D} illustrates the effect of this filter on the laser spectrum for $\alpha=0.7$ and $\beta = 1.3$. These parameters will be used later on for simulations. \\

With this technique, an efficient flattening of the spatial intensity profile is possible, provided that SC is large enough to ensure a good coupling between spatial and spectral degrees of freedom. In practise, we found out that an efficient flattening is possible for a PFT parameter $\xi > \xi_{0}$. Figure \ref{metrs} illustrates the effect of such filtering on the laser spatio-spectral profile at focus, where we assumed an initially Gaussian laser beam with a PFT $\xi = 1.5 \xi_{0}$ and a beam waist of $w_0 = 3.2 \mu m$. Panels (a) and (b) respectively show the properties of such a beam without and with spectral shaping (applied before focusing). Thanks to SC at focus, the pure spectral filtering applied before focusing [cf. Fig. \ref{newfig1D}] damps the laser intensity mostly around  $x_f = 0$, where the local laser frequency is close to the central frequency $\omega_0$ [cf. Fig. \ref{metrs} (c)]. The effect of this spectral shaping on the laser beam profile at focus is revealed on panels (b) and (d) of Fig. \ref{metrs}. The flattening of the laser beam profile at focus is clear.

\begin{table*}[ht]
\begin{center}
\label{refshaping}
%\begin{tabular}{|c|c|c|c|c|c|c|c|c|c|c|c|c|}
\begin{tabular}{|c*{13}{|>{\centering\arraybackslash}p{3em}}}
\hline
\multirow{2}{*}{\textbf{Technique}} & \multicolumn{8}{c|}{\textbf{Laser}}                & \multicolumn{2}{c|}{\textbf{Plasma}} & \multicolumn{2}{c|}{\textbf{PIC}} \\ \cline{2-13}
      & $a_0$ & $\theta_0$ & $\tau_0$ & $w_0$  & $\xi(\xi_{0})$ & $\alpha$ & $\beta$ & $\frac{\Delta z_\xi}{Z_\xi}$ & L  & $n_0$   & $d$    & ppcell  \\ \hline
Simple Gaussian beam    & \multirow{3}{*}{$30$} & \multirow{3}{*}{$55^\circ$} & \multirow{3}{*}{$16 fs$}& \multirow{3}{*}{$3.2\mu m $} & \multirow{3}{*}{$1.5 $}     & 0& 0 & 0   & \multirow{3}{*}{$\frac{\lambda_0}{20}$}   & \multirow{3}{*}{$220 n_c$}  & \multirow{3}{*}{$\frac{\lambda_0}{285}$}  & \multirow{3}{*}{$6$}    \\ \cline{1-1} \cline{7-9} 
Laser defocusing   &  & & & & & 0 & 0 & $-0.31$  &  &   &   &   \\ \cline{1-1}\cline{7-9} 
Intensity shaping    &  & & & & & 0.7 & $1.3$ & 0   &   &   &  &   \\ \hline

\end{tabular}
\end{center}
\caption{ Physical and numerical setups for simulations used to compare the laser pulse shaping technique with the defocusing technique and the standard lighthouse scheme. $d$ is the size of the mesh step along all directions, $ppcell$ the number of pseudo-particles per cell and $n_0$ the bulk PM density (in units of critical density $n_c$). N.B: $a_0$ and $\tau_0$ are laser amplitude and laser duration without $WFR$.  } 
\label{refshaping}
\end{table*}

%Panel (b) sketches three line-outs of the spatial beam profile represented on panel (a) at three different transverse positions : $x = 0.75 w_{\xi}$ (blue curve), $x = 0 $ (green curve), $x =-0.75 w_{\xi}$ (red curve). The black dashed line represents the filter gain function as a function of frequency $\omega$. The laser spectrum at $x=0$ (in green) is significantly damped by the filter such that the corresponding total laser amplitude for adjacent transverse positions is roughly constant (see red curve on panel (d)). comment : à voir si on garde, si modification de la figure 9.

 In the following, we assume that the filtering operation conserves the incident beam energy (This could be achieved using a Dazzler where the energy of the central frequency would be redistributed to other frequencies of the pulse). In these conditions, the modification of the laser spectrum by the spectrum only leads to a modest decrease of the maximum laser intensity at focus by $30\%$ only, compared to the case with WFR only.  In the next subsection, we conduct 2D PIC simulations using such laser beam profile in order to assess the effectiveness of this technique in producing low divergence and angularly separated attosecond light pulses, and compare it with the technique of the previous section.

\subsection{PIC simulations of Doppler harmonic generation with a spatially-flattened laser beam}

The effect of the spatial flattening of the laser beam profile on the separation of attosecond light pulses has been investigated using 2D PIC simulations with the WARP+PXR code. For this matter, we performed three 2D PIC simulations whose parameters are summarized in Table. \ref{refshaping}:

\begin{enumerate}[label=(\roman*)]
\item \textit{Case 1} has been performed employing a standard Gaussian laser beam with WFR to assess the angular separation of Doppler harmonic beams with the lighthouse effect without any tailoring of the laser spatial phase or beam profile. \\
\item \textit{Case 2} has been performed with the same parameters as Case 1, but now using the defocusing scheme to optimize the curvature of the laser spatial phase and reduce harmonic beam divergence. \\
\item Finally, \textit{case 3} has been performed with same parameters as Case 1, but using the spatial flattening of the laser beam profile to reduce harmonic beam divergence.
\end{enumerate}
Note that the same PFT parameter ($\xi=1.5 \xi_0$) has been used in the three cases. 
%\begin{figure}[h]
% \centering
%\includegraphics[width=\linewidth]{figures/new_merged_angulars_shaping.eps}
%\caption{ Integrated angular spectrum (between harmonic orders 15-20) for each simulation case of table \ref{refshaping}.}
%\label{comparewotechniques}
%\end{figure}

\begin{figure}[h]
 \centering
\includegraphics[width=\linewidth]{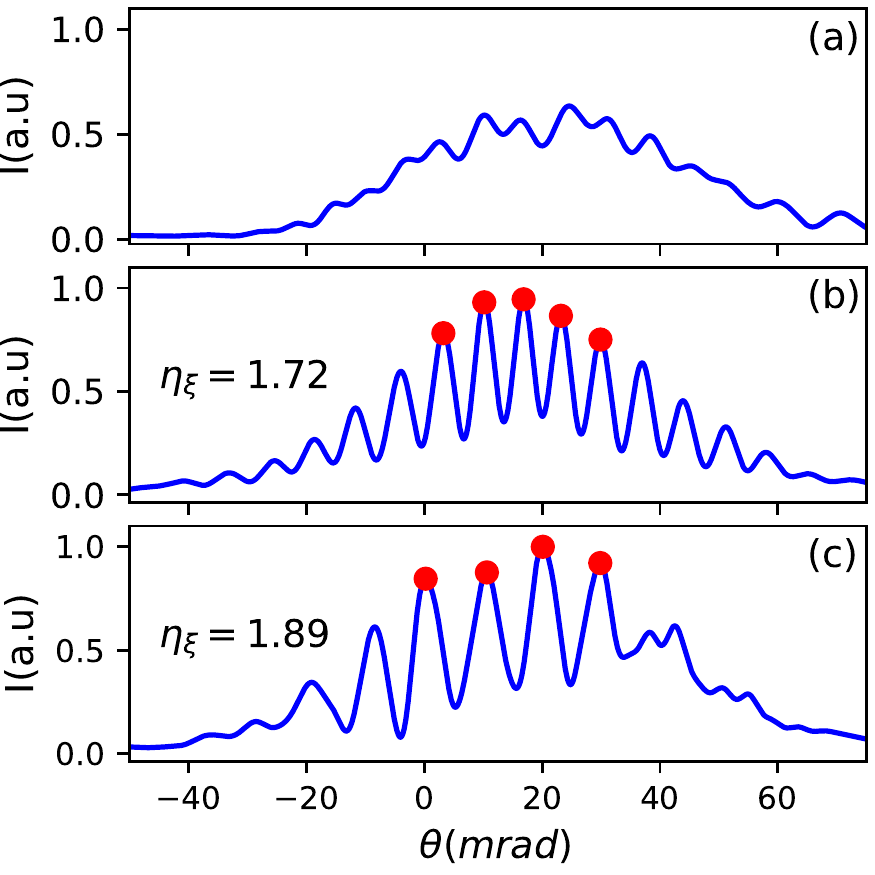}
\caption{Angular profile of the Doppler harmonic beam (between harmonic orders 15-20) obtained for each simulation case of table \ref{refshaping}. Panels (a) to (c) correspond to simulation cases 1 to 3 (see text and table \ref{refshaping}), respectively. On panels (b) and (c), the red dots highlight the direction of emission of the successive attosecond pulses of the train. }
\label{comparewotechniques}
\end{figure}

Figure \ref{comparewotechniques} displays the angular profile of the generated harmonic beams (harmonic orders 15-20) obtained from each of these simulations. As expected, a standard Gaussian beam with WFR cannot produce angularly separated light pulses in the relativistic regime (panel (a)) due to the large harmonic divergence. 

In contrast, panels (b) and (c) show that a good angular separation is obtained using either the defocusing technique or the spatial-flattening one.  The comparison of the results of \textit{Case 2} and \textit{Case 3} is very instructive:
\begin{enumerate}
    \item The angular separation between successive attosecond pulses is larger in \textit{Case 3} than in \textit{Case 2}. This is because the WFR is larger when the PM is placed at the laser best focus (\textit{Case 3}) than when it is slightly out of focus (\textit{Case 2}).
    \item The divergence of the individual attosecond pulses is larger in \textit{Case 3} than in \textit{Case 2}. This is because the focusing effect of the PM cannot be completely mitigated by the shaping of the intensity profile (\textit{Case 3}), while it is fully compensated by adjusting the laser wavefront curvature (\textit{Case 2}).  
    \item A larger number of attosecond pulses are generated in \textit{Case 2} than in \textit{Case 3}. This is because the laser pulse is locally chirped out of focus for \textit{Case 2} [see Eq.(\ref{extdz}], while it is locally Fourier-transform limited at focus for \textit{Case 3}. 
\end{enumerate}

Quantitatively, computing the separation ratio obtained in \textit{case 2} and \textit{case 3} respectively yields $\eta_{\xi}=1.72$ and $\eta_{\xi} = 1.89$, thus showing a comparable angular separation quality between the two techniques. 

\section{Application to a PW-class laser}

\begin{table*}[ht]
\begin{center}
\begin{tabular}{|c*{12}{|>{\centering\arraybackslash}p{4.2em}}}
\hline
\multicolumn{10}{|c|}{\textbf{Physical parameters} }                      & \multicolumn{2}{c|}{\multirow{2}{*}{\textbf{PIC parameters}}} \\ \cline{1-10}
\multicolumn{8}{|c|}{Laser}               & \multicolumn{2}{c|}{Plasma} & \multicolumn{2}{c|}{}       \\ \hline
$a_0$ & $ \theta_0$ &$\tau_0$ & $w_0$  & \multicolumn{3}{c|}{$\xi(\xi_{0})$} & $\Delta z_\xi(Z_\xi)$ & L     & $n_0$  & $d$       & ppcell    \\ \hline
40 & $45^\circ$ & $22 fs$ & $2.4\mu m $ & \multicolumn{3}{c|}{$1$}    & -0.5  & $\frac{\lambda_0}{15}$ & $220 n_c$ & $\frac{\lambda_0}{190}$    & 1     \\ \hline
\end{tabular}
\end{center}
\caption{Physical/numerical parameters for the 3D-PIC lighthouse simulation. $d$ is the step of the spatial mesh along all directions, $ppcell$ the number of pseudo-particles per cell and $n_0$ the bulk PM density (in units of critical density $n_c$). N.B: $a_0$ and $\tau_0$ are laser amplitude and laser duration without $WFR$ and at best focus.  }
\label{3dtable}
\end{table*}

\begin{figure*}[ht]
\centering
\includegraphics[width=\linewidth]{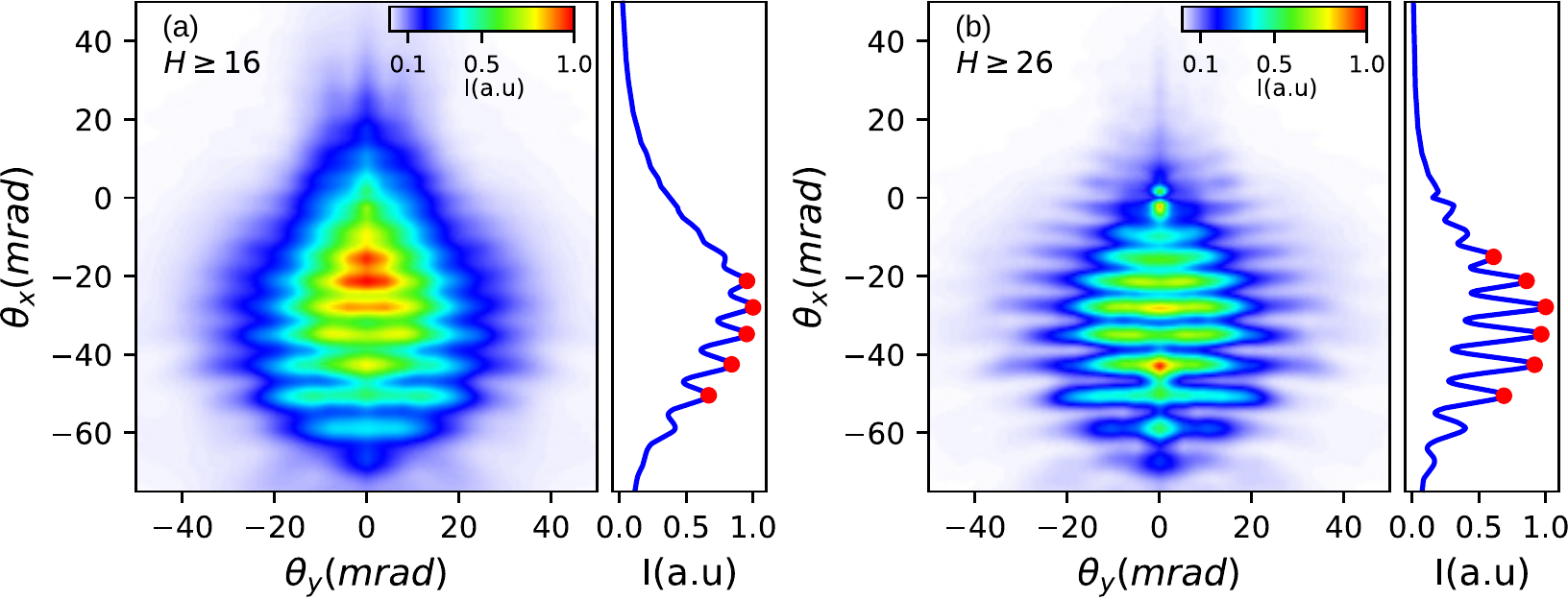}
\caption{Angular profiles of Doppler harmonics obtained from 3D PIC simulations for (a)  harmonic orders $\geq 16$ and (b)  harmonic orders $\geq 26$. The side graphs represent lineouts (integrated along $\theta_y$) of the harmonic angular profiles as a function of $\theta_x$ (direction of WFR). On each panel inset, the red dots highlight the angles of emission of the successive attosecond pulses of the train. }

\label{merged3d}
\end{figure*}

\begin{figure}[h]
 \centering
\includegraphics[width=\linewidth]{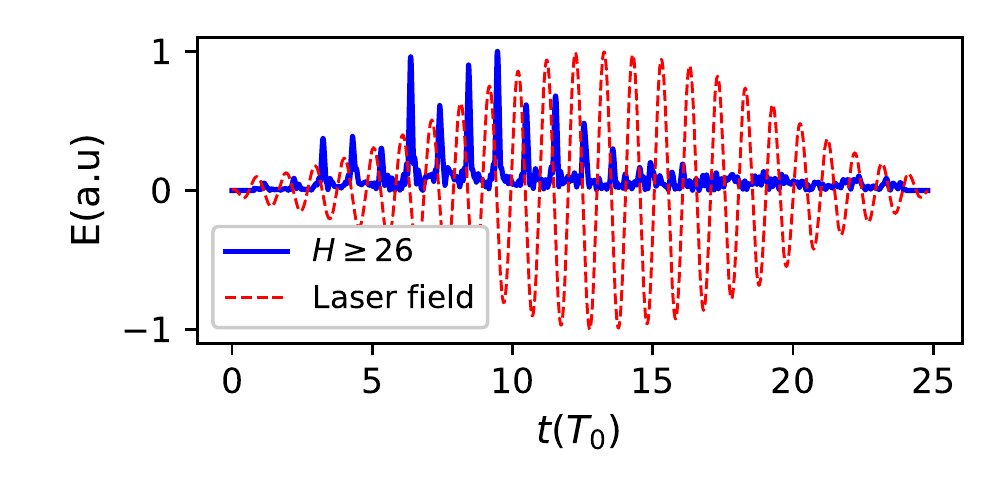}
\caption{Temporal profile of the attosecond pulse train in the 3D simulation.  The blue curve represents the temporal evolution of the reflected field amplitude (where we filtered harmonic orders $\geq 26$). The red curve corresponds to the reflected laser field (fundamental frequency only).}
\label{1dmilieu}
\end{figure}

\begin{figure*}[ht]
\centering
\includegraphics[width=0.9\linewidth]{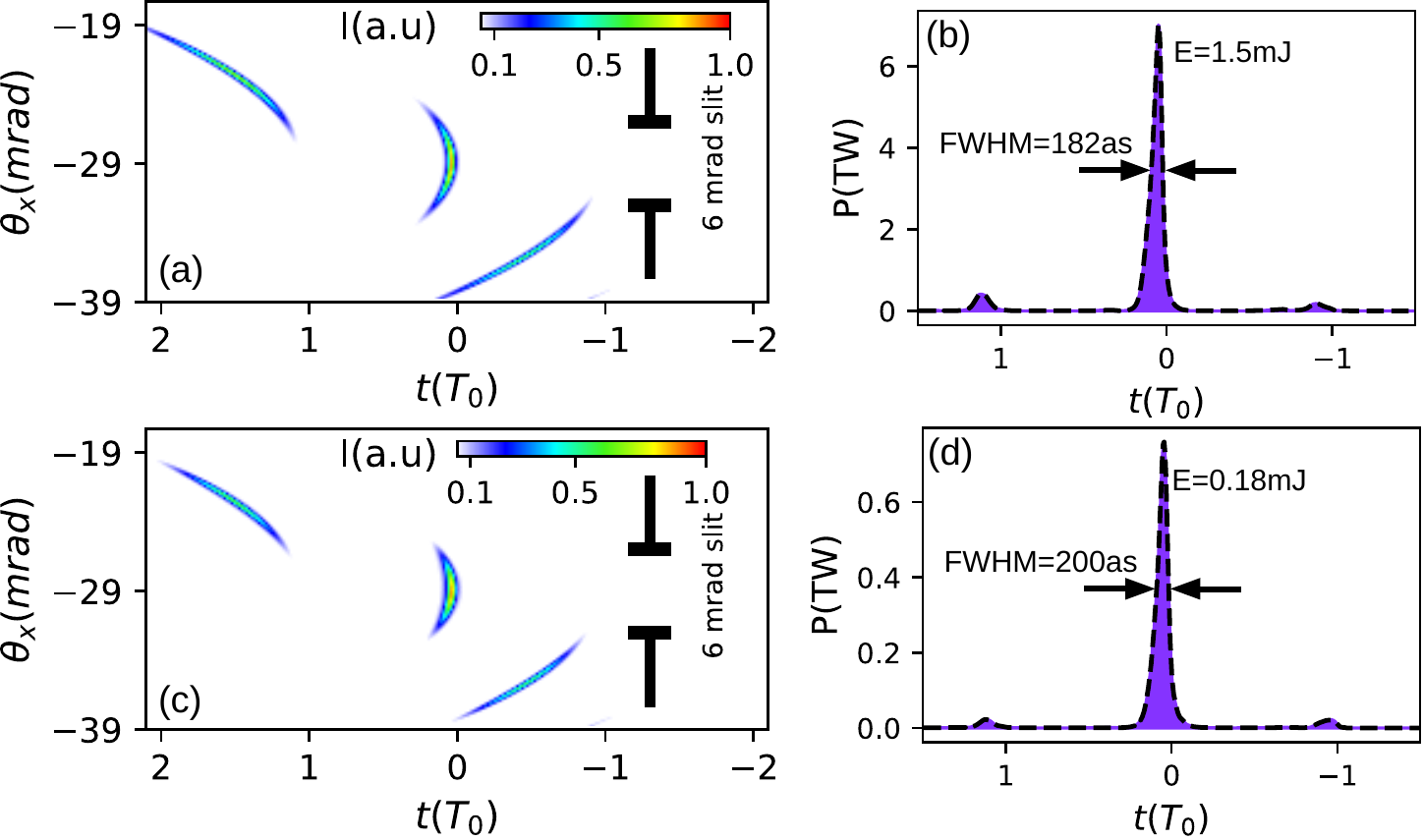}
\caption{Panels (a) and (c): Spatio-temporal profile of central attosecond pulses along the $\theta_x$ axis, for harmonic orders $\geq 16$ (a) and $\geq 26$ (b) . Panels (b) and (d): Intensity profile of the attosecond pulse obtained after the 6 mrad slit filter (sketched in panel (a) and (c)). After the slit, we obtain a central attosecond pulse with a duration of $182$ as and $200$ as (FWHM) respectively and an energy of $1.5mJ$ and $0.18mJ$ leading to a peak power of $7$TW for the first harmonic range (harmonic orders $>16$) and of $0.7TW$ for the second one  (harmonic orders $>26$).}
\label{figure3Dslit}
\end{figure*}

In this section, we use 3D PIC simulations to determine the properties of the attosecond pulses that can be generated with a PW-class laser by combining the attosecond lighthouse effect with the defocusing scheme described in section III. This 3D study enables us to: 
\begin{enumerate}[label=(\roman*)]
    \item validate the 2D PIC simulations ran in the previous sections to assess the efficiency of this divergence reduction technique, 
    \item obtain quantitative estimates of the properties (divergence, energy, duration) of the generated isolated attosecond pulses, 
    \item investigate 3D spatial properties of the harmonic emission that are not accessible with 2D simulations. 
\end{enumerate}

\subsection{Physical/numerical setup}
For this goal, we performed a single 3D PIC simulation of the lighthouse effect using the pseudo-spectral WARP+PICSAR code. The physical and numerical parameters are summarized in Table \ref{3dtable}. Taking into account typical energy losses between the laser output and the target area, this simulation corresponds to a laser of $\approx 1PW$ peak power just after the compressor. The pulse duration of $\tau_0=22fs$ prior to the application of WFR is characteristic of state-of-the-art PW femtosecond lasers.

In this simulation, the reflected field is captured at each time step on a probe plane located at a position $ z_0 \simeq 10 \lambda_0$ from the target surface. This probe field $E(x,y,z_0,\omega)$ is then used an an input to calculate the  spatial properties of the harmonic beam $E(x,y,z,\omega)$ at any arbitrary plane $z$, using plane-wave decomposition. This spectrally-resolved 3D propagation post-processing was computationally very demanding as the reflected field data occupies hundreds of gigabytes of memory. We therefore had to develop specific parallel post-processing tools to parallelize all the distributed Fourier transforms required in the plane wave decomposition method.

This 3D PIC simulation required $32,768$ BLUE GENE-Q nodes of the MIRA supercomputer at ALCF during $20$ hours, leading to a total of $10$ millions core hours for the entire simulation.  \\

\subsection{Simulation results}

Fig. \ref{merged3d} represents the angular profiles of the emitted harmonic beams. WFR occurs along the $x$ axis. The profiles plotted in the side panels are obtained by integrating the 2D angular profile along the direction $\theta_y$.

 From the 1D angular profiles along $\theta_x$, one can estimate that $\eta_\xi=1.15$ for harmonic orders $>16$ (panel(a)) and $\eta_\xi=1.75$  for harmonic orders $>26$ (panel (b)). This shows that the attosecond light pulses are angularly well separated over a large harmonic range. It also validates the efficiency of the defocusing technique in reducing the harmonic beam divergence and achieving angular separation of the successive attosecond pulses of the train with the lighthouse effect.  

On both panels, Figure \ref{merged3d} shows that the harmonic divergence is larger along the $\theta_y$ direction (orthogonal to the direction of WFR) than along the $\theta_x$ direction. In other words, the attosecond light pulses are elliptically shaped in the far-field, with a larger divergence along the axis without WFR. This result from the combination of two effects: 
\begin{enumerate}[label=(\roman*)]
\item In the presence of WFR, the laser focal spot is elliptical on target, with a larger waist along the axis of WFR (i.e. $x$-axis). A larger waist results in a smaller divergence in the far field. 
\item the second effect comes from the different impact of laser defocusing in the two planes. Indeed, the laser is defocused by $\Delta z_\xi$  from the PM surface so that the laser wavefront curvature compensates the PM curvature in the plane of WFR, leading to harmonic divergences close to its diffraction-limited value in this plane. However, in the orthogonal direction, the PM curvature is larger, due to the smaller value of the laser waist. The PM curvature is not fully compensated by laser curvature in this plane, thus leading to a higher harmonic beam divergence. 
\end{enumerate}

A striking feature of Fig.\ref{merged3d} is that there is little to no attosecond light beams emitted for $\theta_{x}>0$. Since emission time is encoded in emission direction in the lighthouse scheme, this suggests that that the emission of attosecond pulses is almost suppressed in the second half of the laser pulse. Fig. \ref{1dmilieu} verifies that this is indeed the case, by  representing the temporal profile of the attosecond pulse train close to the target (blue line). The driving laser field is shown in red as a reference. The harmonic emission efficiency is indeed observed to significantly drop during the second half of the laser pulse. This drop of harmonic efficiency comes from a sharpening of the PM density gradient due to laser radiation pressure (hole boring) \cite{wilks1992absorption,Vincenti2014}. This effect strongly reduces the harmonic generation efficiency (which is higher for longer PM scale length) \cite{chopinaeauprx}. As initially suggested in \cite{alhhenri}, this simulation thus illustrates how the attosecond lighthouse effect can be exploited as a powerful time-resolved probe of the laser-plasma interaction dynamics in experiments.

Finally, Fig. \ref{figure3Dslit} displays the spatio-temporal profiles of attosecond light pulses propagating around the angular position $-27$ mrad, for two different harmonic ranges:  harmonic orders $n>16$ [panel (a)] and harmonic orders $n>26$ [panel (b)]. Each attosecond pulse divergence is as low as $4.5$mrad. Panels (c) and (d) represent the signal obtained after spatial filtering by a 6 mrad slit placed on the path of the central attosecond pulse from panels (a) and (b). The filtered signal is made of one central attosecond pulse, and two satellite pulses stemming from the neighboring attosecond pulses. For harmonic orders $n>16$, panel (b) shows that the central attosecond pulse has a $182$-as duration and carries an energy of 1.5mJ, corresponding to a peak power of $10$TW. The energy contrast of the filtered attosecond signal is larger than $10$. This shows that bright isolated attosecond pulses of $10$TW power (in the 20-50 eV photon energy range) can effectively be obtained with this setup.

\section{Conclusion}

This article proposes two novel techniques, readily applicable in experiments, to reduce the divergence of Doppler harmonics generated on relativistic plasma mirrors and achieve angular separation of the associated attosecond pulses by the attosecond lighthouse effect. The first technique consists in optimizing the curvature of the incident laser wavefronts to compensate for the PM curvature induced by laser radiation pressure and that tends to increase Doppler harmonic divergence. In practice, this is achieved by moving the PM surface slightly away from the laser best focus. The second technique is based on the flattening of the laser beam  intensity profile at focus, to suppress or reduce the laser-induced PM curvature. In the attosecond lighthouse scheme, this is possible by applying a simple spectral shaping to the laser beam. Both techniques have been validated using state-of-the-art 2D and 3D Particle-In-Cell simulations and show excellent angular separation of attosecond light pulses with the lighthouse effect in realistic conditions, using laser pulses with durations of the order of 8 optical periods. This work provides realistic pathways to achieve the lighthouse effect in future experiments with high-power lasers. 

\begin{acknowledgments}
An award of computer time (PICSSAR\_INCITE) was provided by the Innovative and Novel Computational Impact on Theory and Experiment (INCITE) program. This research used resources of the Argonne Leadership Computing Facility, which is a DOE Office of Science User Facility supported under Contract DE-AC02-06CH11357. This work was supported by the French National Research Agency (ANR) T-ERC program (grant: PLASM-ON-CHIP). We acknowledge the financial support of the Cross-Disciplinary Program on Numerical Simulation of CEA, the French Alternative Energies and Atomic Energy Commission. 
\end{acknowledgments}

 \appendix
 
\section*{Appendix}
 \subsection{Analytical form of a Gaussian beam with WFR out of laser focus}
In this appendix, we first derive the analytical expression at an arbitrary position $\Delta z$ of a Gaussian beam with SC at focus (located at $\Delta z=0$) using a plane wave decomposition. 

Fourier transforming Eq. \ref{Extz0} with respect to transverse spatial coordinates ($x$,$y$) and time $t$: 

\begin{multline}
\widehat E(k_x,k_y,\omega, z=0) = \\ \frac{1}{2\pi}\int_{-\infty}^{\infty}\int_{-\infty}^{\infty}\int_{-\infty}^{\infty}  E(x,y,t, z=0)e^{-i k_x x -i k_y y - i \omega t}   \mathrm{d}t  \mathrm{d}x\mathrm{d}y
\end{multline}

yields: 

\begin{multline}
\widehat E(k_x,k_y,\omega, z=0)\propto e^{-\dfrac{w_0^2}{4}k_y^2} e^{-\dfrac{w_\xi^2}{4+\tau_\xi^2w_\xi^2\zeta^2}k_x^2}  \\
e^{-\dfrac{\tau_\xi^2}{4+\tau_\xi^2w_\xi^2\zeta^2}(\omega-\omega_0)^2}e^{-i\dfrac{\tau_\xi^2w_\xi^2\zeta}{4+\tau_\xi^2w_\xi^2\zeta^2}k_x(\omega-\omega_0)}
\label{apppp1}
\end{multline}

In Fourier space, the propagation of the field by $\Delta z$ simply writes: 

\begin{equation}
\widehat  E(k_x,\omega,z=\Delta z)  = \widehat E(k_x,\omega,z=0)  e^{ik_z \Delta z}
\label{propdz}
\end{equation}

where:
\begin{equation}
k_z = \sqrt{\frac{\omega^2}{c^2}-k_x^2-k_y^2}
\label{app2}
\end{equation}

Under the paraxial approximation $k_x\ll k =\omega/c$ and $k_y\ll k =\omega/c$, the above equation becomes: 

\begin{equation}
k_z \approx \frac{\omega}{c}-\frac{k_x^2}{2k}-\frac{k_y^2}{2k}
\label{app2}
\end{equation}

 In the following, we assume that $k=k_0+\Delta k\approx k_0$ in the second term of the right hand side of Eq.\ref{app2}. Physically, this approximation means that all the frequency components of the laser beam diffract the same way. This holds as long as $\Delta k/k_0\ll 1$, which is verified for a laser pulse duration of a at least a few optical cycles. As introduced in section \ref{currentlimsub}, high-power lasers considered here are at least $5$ optical cycles long and we can thus reliably use:

 \begin{equation}
k_z \approx \frac{\omega}{c}-\frac{k_x^2}{2k_0}-\frac{k_y^2}{2k_0}
\label{app2}
\end{equation}

Note that for a laser pulse duration close to a single optical cycle, other spatio-temporal couplings may arise from a different diffraction of the different frequency components and the above approximation fails. 

Fourier transforming back Eq. \ref{propdz} along $k_x$, $k_y$ and $\omega$ : 

\begin{multline}
\label{bigequationfgsdfjq}
E(x,y,t'=t-z/c,z=\Delta z) = \\ \frac{1}{2\pi}\int_{-\infty}^{\infty}\int_{-\infty}^{\infty}\int_{-\infty}^{\infty}   \widehat  E(k_x,\omega,z=\Delta z) \mathrm{d}k_x \mathrm{d}k_y \mathrm{d}\omega
\end{multline}

finally yields: 

\begin{multline}
\label{extdz_appendix}
E(x,y,t'=t-\Delta z/c,z=\Delta z)   \\  \propto e^{- \dfrac{k_0 y^2}{2 \big (Z_R+i\Delta z\big)}} e^{- \dfrac{k_0 x^2}{2 \big (Z_\xi+i\Delta z\big)}} \times e^{-\dfrac{t'^2}{\tau^2_{\xi}}  \left[1+\dfrac{(\xi/\xi_0)^2 \Delta z^2}{\Delta z^2+Z_\xi^2}\right]} \\ 
 \times e^{ -\dfrac{i(\xi/\xi_0)^2t'^2}{\tau^2_\xi\left(Z_\xi/\Delta z+\Delta z/Z_\xi\right)}} \times e^{i\left[\omega_0 + \dfrac{\zeta x}{1+i \dfrac{ \Delta z}{Z_\xi } }\right]t'}
\end{multline}
where $Z_R=\pi w_0^2/\lambda_0$ is the laser Rayleigh range without SC and $Z_{\xi}=\pi w_\xi^2/\lambda_0$ is the laser Rayleigh range in the plane of SC.

Let us analyse the physical meaning of each term in Eq. \ref{extdz_appendix}: 
\begin{enumerate}[label=(\roman*)]
    \item The first two terms correspond to the spatial amplitude and phase profiles of the laser at $\Delta z$ along $y$ and $x$ directions. From this term we can deduce the following formulae for the laser waist $w_{\xi}(\Delta z)$ at $\Delta z$ in the plane of SC: 
    \begin{equation}
    w_\xi(\Delta z) = w_\xi\sqrt{1+\left(\frac{\Delta z}{Z_\xi}\right)^2}
    \end{equation}
    as well as the laser wavefront radius of curvature $R_{\xi}(\Delta z)$: 
    \begin{equation}
    R_\xi(\Delta z) = \Delta z + \frac{Z_{\xi}^2}{\Delta z}
    \end{equation}
    \item the second term corresponds to the local laser temporal profile, from which we can deduce the modified laser pulse duration at $\Delta z$: 
    \begin{equation}
        \tau_\xi(\Delta z) = \tau_\xi\sqrt{\dfrac{1+(Z_\xi/\Delta z)^2}{1+(\xi/\xi_0)^2+(Z_\xi/\Delta z)^2}}
    \end{equation}
    As one moves closer to the best focus $z\rightarrow 0$, we find $\tau_\xi(z) \rightarrow \tau_\xi$, while in the far field  $z\rightarrow\infty$, $\tau_\xi(z)\rightarrow\tau_\xi/\sqrt{1+(\xi/\xi_0)^2}=\tau_0$, as expected.
    \item the third term is a phase term corresponding to a temporal chirp $\beta_\xi(\Delta z)$ given by : 
    \begin{equation}
        \beta_\xi(\Delta z) = -\dfrac{(\xi/\xi_0)^2}{\tau_\xi^2\left(\frac{Z_\xi}{\Delta z}+\frac{\Delta z}{Z_\xi}\right)}
    \end{equation}
    This term tends to zero at best focus.
    \item the last term corresponds to a mix of WFR and PFT term. The PFT term $\xi(\Delta z)$ is given by its real part: \begin{equation}
        \xi(\Delta z) = \dfrac{\zeta}{\frac{\Delta z}{Z_\xi}+\frac{Z_\xi }{\Delta z}}
    \end{equation}
    When $\Delta z\rightarrow \infty$ one finds that $\xi(\Delta z)\rightarrow 0$. This is expected as the beam waist $w_{\xi}(\Delta z)\rightarrow\infty$ while the pulse duration $\tau_\xi(\Delta z)\rightarrow\tau_0$. However, one can check that $\xi(\Delta z)w_{\xi}(\Delta z)\rightarrow \tau_0$ when $\Delta z\rightarrow \infty$ and $\xi$ is initially set such that $\xi w_0 = \tau_0$ before focusing. The SC $\zeta(\Delta z)$ is given by the imaginary part of the last term of Eq. \ref{extdz}: 
    \begin{equation}
        \zeta(\Delta z) = \dfrac{\zeta}{1+\left(\frac{\Delta z}{Z_\xi}\right)^2}
    \end{equation} 
\end{enumerate}

 \subsection{Expression of WFR velocity out of laser focus}

Here, we use the expression of the laser field derived in the previous section to deduce an analytical formula for the WFR velocity out of laser focus. The WFR effect is all encoded in the phase of the last exponential. The velocity $v_\xi(\Delta z)$ at a distance $\Delta z$ from focus is given by: 
\begin{equation}
    v_\xi(\Delta z) =\frac{d\Theta(\Delta z,t')}{dt'}
\end{equation}
where $\Theta = k_x/k$, $k_x(\Delta z,t') = \zeta(\Delta z) t' $ and $k(\Delta z,'t) = \omega_0-t'\beta(\Delta z)/c+\zeta(\Delta z) x/c $. In the following, we derive the expression of WFR velocity at the center of the laser beam $x=0$ and neglect the variation of $k$ with $x$ due to spatial chirp. With the attosecond lighthouse effect, we use $\xi\approx \xi_0$ to maximize WFR while avoiding a too high reduction of laser intensity at focus. In addition, as we show in the manuscript, the required defocusing distance is of the order of $Z_{\xi}$. This implies that $\beta\approx 1/\tau^2_\xi$. In these conditions, the laser frequency variation due to temporal chirp is of the order of $\beta \tau_\xi \approx 1/\tau_\xi$ which is negligible compared to $\omega_0$ in the approximation $\tau_\xi\gg T_0$ where $T_0$ is the laser period. As a result, we can assume $k(\Delta z,'t) \approx k_0$, which gives: 
\begin{equation}
    v_\xi(\Delta z) = \dfrac{v_\xi}{1+\left(\frac{\Delta z}{Z_\xi}\right)^2}
    \label{vxi}
\end{equation}

\bibliography{refs}        %use a bibtex bibliography file %refs.bib

\end{document}